# Learning Newtonian mechanics with an intrinsically integrated educational game

Anne van der Linden[a], Ralph F. G. Meulenbroeks[b] and Wouter R. van Joolingen[c]

Freudenthal Institute, Utrecht University, Utrecht, The Netherlands

[a] a.vanderlinden@uu.nl, ORCID: 0000-0001-7970-0664
[b] r.f.g.meulenbroeks@uu.nl, ORCID: 0000-0001-6614-9156
[c] w.r.vanjoolingen@uu.nl, ORCID: 0000-0002-4271-2861

Corresponding author: Anne van der Linden. a.vanderlinden@uu.nl

**Abstract**

**Background**: Research on cognitive effects of educational games in general shows promising results. However, large variations in learning outcomes between individual educational games exists. Research on the design process and different design elements of educational games has led to some interesting directions, but some design aspects remain unclear.

**Objectives**: We examined how an educational game designed on the basis of intrinsic integration theory, based on a strong alignment between game and learning goals, supports the learning of Newtonian mechanics.

**Methods**: This study applied a mixed-methods approach (N=223). An pre- and posttest design was used to examine possible learning and transfer effects fostered by playing the educational game, Newton's Race. To examine how players played the game, log data of each player was digitally recorded during gameplay.

**Results and Conclusions**: Our findings demonstrated a significant positive learning effect of Newton's Race ($p = .003$, $d = .201$). This finding can be explained with acquired log data. Log data show that players' gameplay mostly matched expected learning during the game, with physically correct game settings occurring more and more as gameplay progressed. The ability to transfer learned knowledge onto other situations was shown to be limited to situations closely resembling the game environment.

**Implications**: Similar designed intrinsically integrated games on different (physics) subjects could also foster learning in a relative short time. In order to foster transfer to other situations we propose embedding the game within other instructional activities.

**Key words**: Educational games; Intrinsic integration; Newtonian mechanics; Secondary Education.

1. Introduction

The potential benefits of using digital games for education (i.e., educational games) have not gone unnoticed by educational researchers. Benefits for using an educational game are mainly focused around the concept of flow (Csíkszentmihályi, 1990). In commercial games this flow is evident in the player being fully immersed in the game, forgetting where they are and what time it is. As a result, when players are faced with a problem in the game they want to readily find the solution in order for them to continue the gameplay and keep the sense of flow. This evident problem-solving attitude whilst playing a game could be useful for education. This educational application of a digital game environment can be specifically interesting for some domains, such as physics. While traditional experiments are bound to the actual laws of physics, a digital game allows for physical parameters to deviate from the set parameters on this earth. This creates new learning opportunities as the consequences of these deviating parameters become evident in the game environment.

Research on cognitive effects of educational games in general shows promising results (Clark, Tanner-Smith & Killingsworth, 2016). However, large variations in learning outcomes between individual educational games exists. This could very well align with the large variations that exists within the design of the educational games. Research on the design process and different design elements of educational games has led to some interesting directions, but some design aspects remain unclear (Clark et al., 2016; Denham, 2016; Ke, 2016; Lameras, Arnab, Dunwell, Stewart, Clarke & Petridis, 2017). For instance, increased interest in methodologies for educational game design (Ávila-Pesántez, Rivera & Alban 2017), is rarely reflected in detailed descriptions on the interaction between gameplay and learning objectives. Therefore, many intricacies of combining learning with gameplay remain obscure (Czauderna & Guardiola, 2019). As Zeng, Parks and Shang (2020) stated, "educational game design still has challenges to balance education and gameplay" (p. 193).

## 1.1 Intrinsic integration

A possible way to increase the learning effects of a game may be to integrate the educational content with the gameplay itself (Denham, 2016). This is referred to as intrinsic integration: the subject matter and the game idea are integrated (Kafai, 1996). In an intrinsically integrated game, the game goal is aligned with the learning goal as closely as possible, in the sense that reaching the game goal coincides with reaching the learning goal. This is not easily achieved, however, and several studies detail the difficulties encountered in attempting to integrate learning with gameplay without affecting the enjoyability of the game (Vandercruysse & Elen, 2017; Czauderna & Guardiola, 2019).

In a game the player interacts with the game environment through *game mechanics,* in order to reach a game goal. Sicart (2008) accordingly defines game mechanics as 'methods invoked by agents, designed for interaction with the game state'. Examples of game mechanics are jumping over obstacles, trading with other gamers, and climbing structures in the game environment. When designing an *educational game,* however, the game also becomes a *learning environment*, which requires a design based on a chosen pedagogical approach.

The main tool for integrating a pedagogical approach within a game is to make proper use of the game mechanics. Proper alignment of pedagogical approach and game mechanics can thus support the alignment of game goal and learning goal. This completes the idea of *intrinsic integration*, aligning at both the goal level and the mechanism level which creates a very strong and seamless game-based learning situation. We used a guiding frame for designing our intrinsically integrated game as shown in Figure 1 (Van der Linden, van Joolingen, & Meulenbroeks, 2019). Having this framework does not necessarily make the design of intrinsically integrated games straightforward, however. The focus of the current study is not on the game design process itself, it is on utilizing the guiding frame in order evaluate the designed game. In this study, we studied both potential learning effects of the game as well as the effect of the implementation strategy. Using log data on players' actions during gameplay, we evaluate the extent to which the goal of intrinsic integration was reached.

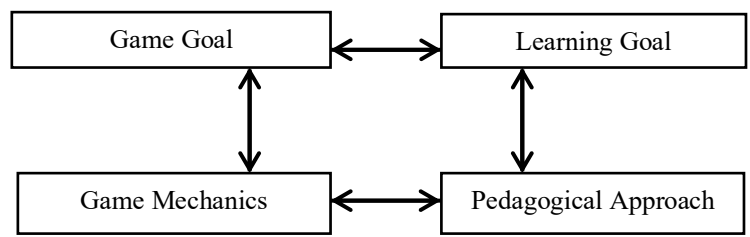

**Figure 1.** Guiding frame on alignment between the game goal, learning goal, pedagogical approach and game mechanics for designing an intrinsically integrated educational game, such as Newton's Race.

**1.2 Implementation of the guiding frame: Newton's Race**

The above guiding frame was implemented in the design of Newton's Race. For the purpose of this study a short description of this implementation, with the main focus on the pedagogical approach, is described below. A description of the gameplay of Newton's Race can be found in the materials and methods section of this paper and a more detailed version of the design process can be found here (Van der Linden, van Joolingen, & Meulenbroeks, 2019).

### 1.2.1 Learning goal

The subject context of Newtonian mechanics is chosen because the conceptual challenges students encounter within this subject have been well-studied. Within this subject students' existing ideas based on daily life experiences, are quite persistent (Halloun & Hestenes, 1985; Fazio & Battaglia, 2018). One of the more persistent conceptual challenges for students concern Newton's second law, stating that force is proportional to acceleration, not to speed (Driver, Squires, Rushworth & Wood-Robinson, 1994). This means that an object can move without forces acting on it and if there is a net force working on an object it cannot remain in a state of constant velocity, but it is accelerating or decelerating. The learning goal that matches with this conceptual challenge is that *students can reason about the effects of forces on different types of motion*. This implies that students understand the effects of forces on motion.

### 1.2.2 Pedagogical Approach

Pedagogical approaches are designed to support students in reaching a learning goal. Before receiving formal instruction in Newtonian mechanics, students have already gathered much daily experience with moving objects. When taught about those movements in a formal scientific context, students must somehow integrate these daily experiences within the scientifically correct theory. In order for students to accomplish this, they need a compelling reason to change their pre-existing ideas. Without an apparent conflict resulting from the application of daily experiences ("naïve theories") to more general and formal situations, it is notoriously difficult to make students understand or even accept the scientific explanations. Students need to be confronted with the fact that their pre-existing ideas and concepts are insufficient in a more formal context (Vosniadou, 1994; Duit & Treagust, 2003; Schumacher, Hofer, Rubin & Stern 2016). In other words, students need to be confronted with the consequences of their expectations and experience so-called cognitive conflict (Hewson & Hewson, 1984) in order for them to abandon or adjust their naïve theories and assimilate the new scientific one.

The pedagogical approach chosen for the game is based on this idea, i.e., the problem posing approach (Klaassen, 1995; Vollebregt, 1998; Kortland, 2001). This approach has been shown to produce promising results on several topics, such as radioactivity and an initial particle model. The problem posing approach demonstrates that students should first see the point of what they are doing during the learning process, before accepting the need for new knowledge and new theories (Lijnse & Klaassen, 2004). The key element of this approach is that students are to be actually confronted with the *consequences* of their pre-existing ideas. When they are confronted with something that counters their expectations, ideally, they should see the need to alter their own theories and adopt the scientific one.

In the context of Newtonian mechanics this confrontation cannot be achieved with experiments in the real word, because we are bound to the way forces actually do work. For example, when riding a bike with a constant velocity in real life, you always actually need to apply force and a car driving at constant velocity does still consume gas or electric energy. So in daily life, constant velocity without exerting force does not exist. In a digital world, however, players can freely test their ideas and experience the effects of those ideas. They can then answer the question: Does this correspond to a real life situation, and adjust their formal ideas accordingly.

### 1.2.3 Game Mechanics

From the problem posing approach two main concepts arise that must be present in the game. First, players must be able to play around with different force options on an object and, secondly, they must subsequently experience the effect of their chosen setting on the object. Thus, in an attempt to align the pedagogical approach with game mechanics, game mechanics are required allowing players to experiment with different force-options and that let players experience the effects of these options.

### 1.2.4 Game goal

Ideally, the game goal and learning goal are aligned in such a way that one cannot be reached without the other. In Newton's Race players have to finish a trajectory with a ball. Players are able to choose different force-options on that ball. However, the trajectory is constructed in such a way that players should only be able to reach the finish line when making a scientifically accurate motion. If players do so, the learning goal is reached and only then can they reach the game goal, the game and learning goal are thus aligned.

## 1.3 Research questions

Studies on previous versions of Newton's Race focused only on the design process and includes a small pilot (Van der Linden & van Joolingen, 2016; Van der Linden, van Joolingen, & Meulenbroeks, 2019). For the current paper a large scale study was conducted to examine how playing Newton's Race actually supports the learning of Newtonian mechanics. We therefore pose the following research question: *How does playing an intrinsically integrated game support learning of Newtonian mechanics?* This was decomposed in the following sub-questions:

SQ1: To what extent does participants' conceptual understanding of Newtonian mechanics change as a consequence of playing the game?

SQ2: To what extent does the acquired knowledge transfer to different situations not closely related to the game?

SQ3: To what extent does the participant's gameplay demonstrate evidence of intrinsic integration according to the guiding frame (Figure 1)?

## 2. Materials and methods

### 2.1 Newton's Race: gameplay

Newton's Race consists of five levels with increasing difficulty. In each level players need to finish a trajectory with a ball. Levels start with a setting phase where players choose between different force-options working on the ball after kicking the ball, see Figure 2. Players are asked to make a realistic motion. The options are: (a) no force working on the ball in the direction of movement (F=0N), which is the scientifically correct setting, (b) a force working on the ball equal to the friction of the surface (F=friction) or (c) a force working on the ball which is bigger than friction (F>friction). After selecting an option, the ball starts moving and the player must guide the ball to the finish line of the trajectory. Depending on the chosen setting the ball will: (a) decelerate, (b) move with a constant velocity or (c) accelerate.

Players can thus explore their pre-existing ideas about forces and motion. For instance, players may choose setting (c) F>friction. In the game this results in a ball forever accelerating. Ideally, students will recognize that in a real world scenario the ball will not accelerate forever, it will decelerate. Players are confronted with something that counters their expectations:, they expect a deceleration, but they experience an acceleration. Then, again ideally, they will come to the realization that their current idea does not work and they will accept the need for a new theory.

They can try different force-options, until they find the one that corresponds with a real life situation, option (a).

This can only work if players experience the effects of their chosen setting, therefore it is important that the resulting type of motion (acceleration, deceleration, constant velocity) is properly visualized. A speedometer, accelerometer, and a graphic comic-book like "tail" emanating from the ball and proportionate to the ball's velocity, are added to the game, see Figure 2.

On the way to the finish line, players can collect coins and accumulate points. The coins are placed in hard to reach places, such as the inside of turns, to make the game more challenging for more experienced gamers. Players can change the direction of the ball by applying a force perpendicular to the direction of motion.

Levels are constructed in such a way that players should only be able to reach the finish line with the correct scientific setting (F=0N). With the other settings the speed of the ball is larger, or even always increasing, making it nearly impossible to get through certain turns in the trajectory. This results in the ball falling off the track and not reaching the finish line. Therefore, the game goal can only be reached with the correct scientific setting. This results in the need for players to make a realistic movement with the correct scientific setting (F=0N). If players do so, the learning goal is reached and only then can they reach the game goal, the game and learning goal are thus aligned.

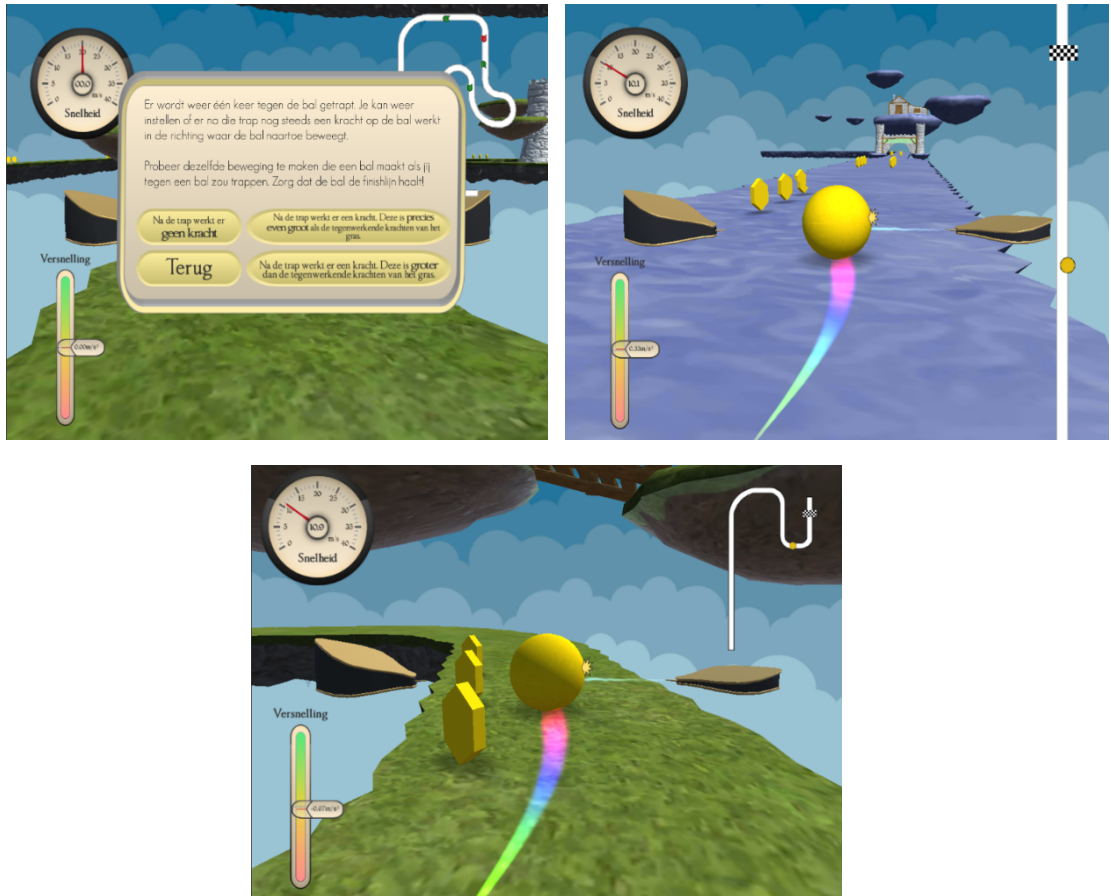

**Figure 2.** Three screenshots of the game. The setting phase (top left), gameplay of level 1 where there is no friction (top right) and gameplay of level 3 (bottom) with friction.

## 2.2 Research design

An pre- and posttest design was used to examine possible learning and transfer effects fostered by playing Newton's Race (SQ1 and SQ2). For each player log data was digitally recorded during gameplay to examine gameplay (SQ3). In addition to answering the research questions, data on different design elements was collected for recommendations on improving the game's visual interface. Therefore, four different versions of Newton's Race were created for this study, which differed only in visual aspects. An example of the difference in visuals can be found in Figure 3. Players were randomly assigned to one of the versions.

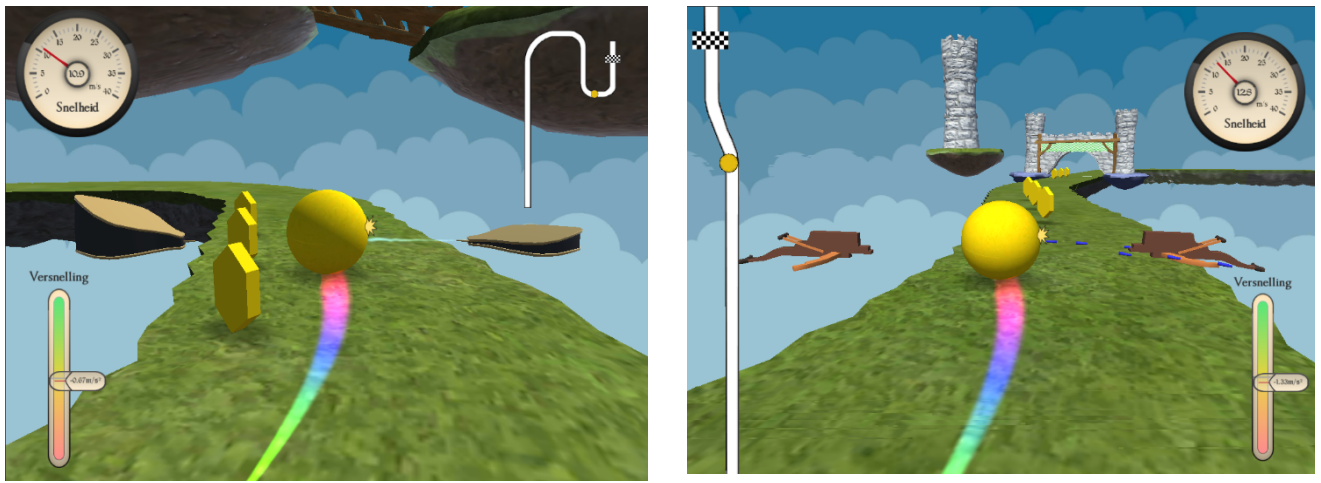

**Figure 3.** Screenshots of two different versions of the game. In addition to these two versions, a version with bellows and the meters on the right side of the screen and a version with the crossbows and the meters on the left side of the screen were used.

## 2.3 Participants

Participants were recruited in a science museum. Children between the ages of 10 and 15 voluntarily participated in this study. Parents provided a written consent at the start of the study. From a total of 223 participants, 117 (52,5%) were boys and 106 (47,5%) were girls. 171 Participants (76,7%) were in primary school and 52 participants (23,3%) were in lower secondary education.

## 2.4 Instruments

The pretest consisted of eight questions, three of which were to assess participants' baseline knowledge, necessary for playing the game. The other five questions were matched to five similar questions in the posttest for determining a possible learning effect. The posttest also contained two questions for assessing knowledge transfer to other situations outside the game. These questions were paired with two similar questions that referred to situations in the game. The questions used in the pre- and posttest are based on questions of the Force Concept Inventory (Hestenes et al., 1992). The posttest included 10 additional questions on the game experience, reflected in the use of for example the speedometer. With these questions, information on the speedometer, accelerometer, the game's difficulty and enjoyability was gathered by using Likert Scales ranging from -2 to 2 points. The pre- and posttest questions were translated from Dutch and can be found in appendix A and B.

## 2.5 Data collection and analysis

The duration of a game session was thirty minutes, including the pre- and posttest. Personal data, such as age and school grade, as well as pre- and posttest questions were administered using a tablet computer. The same tablet was used for playing the game. The intervention time, playing Newton's Race, was fifteen minutes. For each player log data were digitally recorded on attempts, settings and completion of each level. Also observations were made on reactions of participants whilst playing the game.

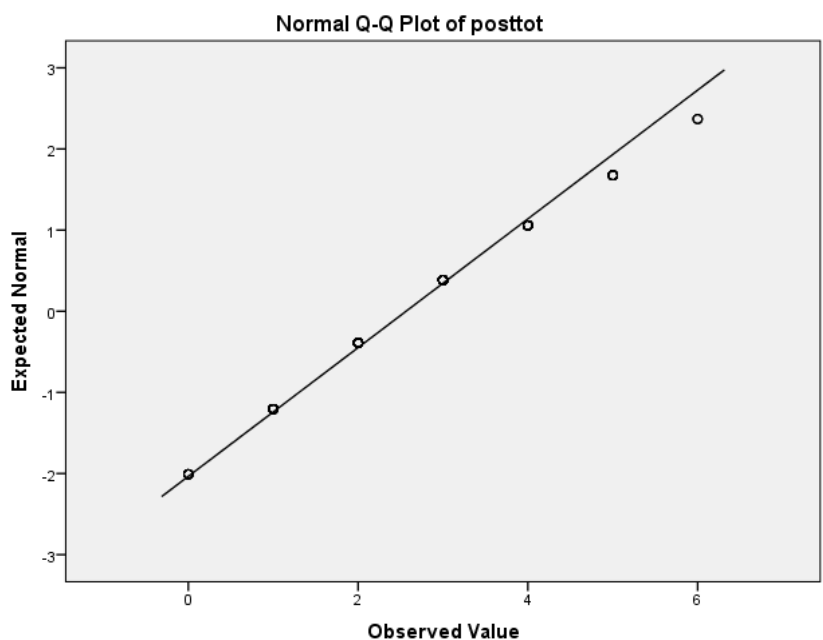

**Figure 4:** The Q-Q plot of the posttest data shows a small deviation on the normal distribution of the upper datapoints.

Q-Q plots of the pretest data and posttest data show a deviation on the normal distribution of the upper datapoints. An example of a Q-Q plot can be found in Figure 4. However, an one-way ANOVA is a relatively robust method against deviations from normality and this method was therefore used to check whether there was a difference in learning effect for the different versions of the game. A paired sample *t*-test was used to determine a possible learning effect between the pre- and posttest questions (SQ1). A further paired sample *t*-test was used to evaluate the transferability of players knowledge to different situations than a ball (SQ2). Effect sizes, Cohens *d*, were also calculated. Furthermore Pearson's correlation was calculated to examine an association between age and learning. With the pretest scores, posttest scores and the log data on the settings of completed levels, the relation was examined between the use of the correct setting in the last completed level and pre- and posttest scores.

From the five levels of Newton's Race, only three levels are directly relevant to the learning goal. Level 1 is an introductory level for players to get used to the game mechanics. In level 4 a new game element is introduced, players get used to this game element in this level. Therefore, only levels 2, 3 and 5 are the key levels, where we expect learning to take place

## 3. Results

### 3.1 General game characteristics

Before the discussing the data related the research questions, we start with giving some general characteristic how the game was played:

In general, players found the game difficult (M = -.45, SD = .994); $t(222) = -6.806$, $p < .000$ and fun (M = .87, SD = 1.068); $t(222) = 12.164$, $p < .000$. Detailed results will be shown following the sub-questions.

For the current research questions we considered the differences between the versions irrelevant, since there was no statistical significant difference between the versions of the game for pretest scores as determined by a one-way ANOVA ($F(3,219) = .125$, $p = .945$). There was also no significant difference between the versions for posttest scores ($F(3,219) = .950$, $p = .417$) either.

### 3.2 SQ1: Conceptual understanding

The first sub-question was: To what extent does participants' conceptual understanding of Newtonian mechanics change as a consequence of playing the game?

**Table 1:** Results of the pre- and posttest (with a minimal value of 0 and a maximal value of 5).

|          | Mean | SD     |
|----------|------|--------|
| Pretest  | 1.66 | .8218  |
| Posttest | 1.87 | 1.0132 |

A paired samples *t*-test was performed to examine the mean differences between the pretests (M = 1.66, SD = .821) and the posttests (M = 1.87, SD = 1.013). A significant difference was found; t(222) = 3.000, *p* = .003; *d* = .201. Results of the Pearson correlation, which tested correlation between age and pretest scores, indicated that there was no significant relation between players' age and the pretest scores (r(222)= .016, *p* = .808). However, another Pearson correlation indicated that there was a significant positive relation between players age and the posttest scores (r(222)= .192, *p* = .004). This may be interpreted as implying that the older the student is, the better the game fosters learning.

### 3.3 SQ2: Transfer

The second sub-question was: To what extent does the acquired knowledge transfer to different situations not closely related to the game?

**Table 2:** Results of the questions closely related to the game and questions about different situations (with a minimal value of 0 and a maximal value of 3).

|           | Mean  | SD   |
|-----------|-------|------|
| Game      | 1.099 | .569 |
| Different | .700  | .611 |

On the questions for knowledge transfer we found that participants scored significantly lower on the transfer questions (M = .700, SD = .611) than on the similar questions on game situations (M = 1.099, SD = .569); t(222) = 7.865, *p* < .001; *d* = .523. Results of the Pearson correlation again indicated that there was a significant positive association between players age and the posttest transfer scores (r(222)= .177, *p* = .008).

### 3.4 SQ3: Gameplay

The final sub-question was: To what extent does the participant's gameplay demonstrate evidence of intrinsic integration according to the guiding frame?

In order for players to understand the game mechanics and the setting phase in each level, players require a certain level of baseline knowledge. An average score was 2.70 out of a maximum score of 3.00 (SD = .522) was measured. On the basis of this, the baseline knowledge was considered adequate. Results of the Pearson correlation, which tested correlation between age and scores indicated that there was no significant relation between players age and their baseline knowledge (r(222)= .185, *p* = .006).

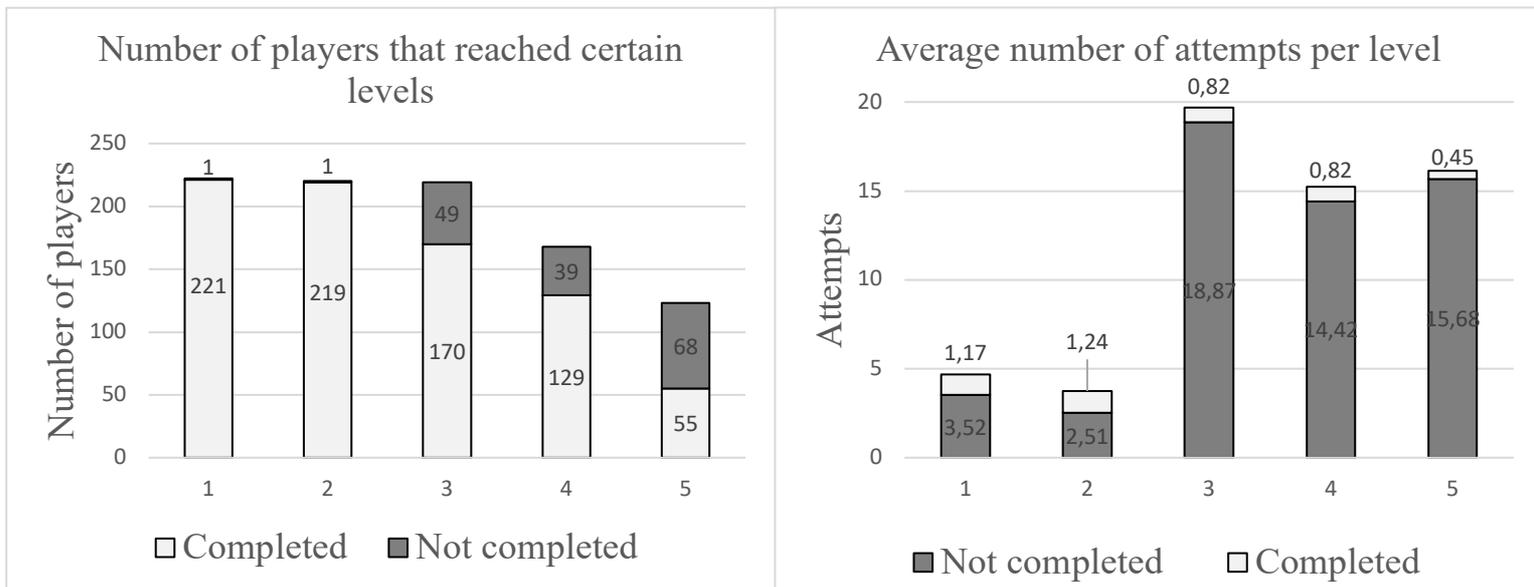

**Figure 5:** In the left graph the total number of players who played each level is displayed. Per level the number of players who completed or not completed the level is indicated. In the right graph the average number of attempts per player for each level is displayed. Also, the average number of attempts from players who completed or not completed the level is indicated.

From the log data information was derived on how players played the game: what settings were chosen during each level. Figure 5 gives an overview on the number of players who succeeded each level and the average number of attempts per player per level. 100 Players completed all the levels and level 3 took the most attempts.

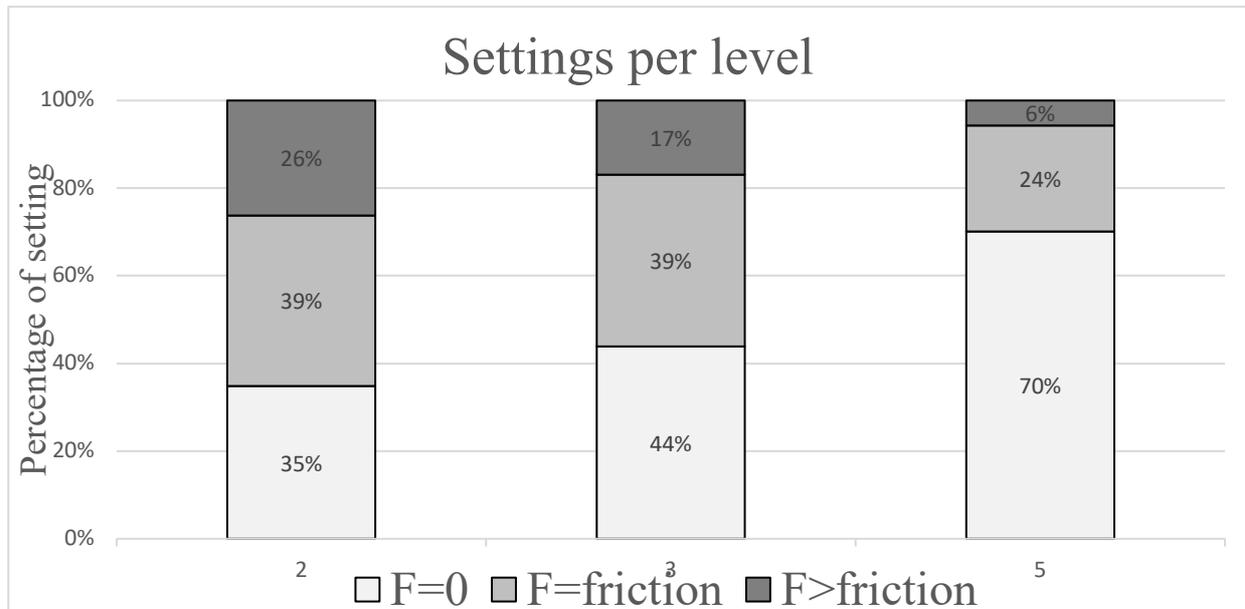

**Figure 6:** For each key level the percentage of the three different chosen settings, F is equal to 0N (correct setting), F is equal to friction and F is bigger than friction, is shown. For these percentages all attempts (completed and not completed) are taken into account.

The percentage of chosen settings of the key levels (levels 2,3 and 5) are shown in Figure 6. To determine whether there are significant differences in the number of the chosen settings per key level, Chi-square tests were performed. Chi-square tests determine the differences between the settings for each level to be significant. Level 2: $\chi^2(2, N = 222) = 20.900$, $p <.000$. Level 3: $\chi^2(2, N = 222) = 537.121$, $p <.000$. Level 5: $\chi^2(2, N = 222) = 1310.535$, $p <.000$. This means that there are significant differences in the number of chosen settings for each key level. Further Chi-squared tests were used to explore these differences, the results of these tests can be found in Table 3. A significant shift in chosen settings is found. In all three levels the setting F=0N, the scientifically correct setting, is chosen significantly more often than the setting F>friction. Also, a significant shift in the settings is found. In level 2 there is no significant difference between the setting F=0N (correct setting) and F=friction. In level 3 and 5, F=0N is chosen significantly more often than F=friction.

**Table 3:** Chi-square tests results between the number of different settings (F is equal to 0N (correct setting), F is equal to friction and F is bigger than friction) per key level.

|  | Level 2 | | Level 3 | | Level 5 | |
|---|---|---|---|---|---|---|
|  | $\chi^2$ | $p$ | $\chi^2$ | $p$ | $\chi^2$ | $p$ |
| F=0N & F=friction | 1.901 | .168 | 11.729 | .001 | 442.361 | .000 |
| F=0N & F>friction | 10.022 | .002 | 516.348 | .000 | 1088.384 | .000 |
| F=friction & F>friction | 20.531 | .000 | 380.191 | .000 | 229.614 | .000 |

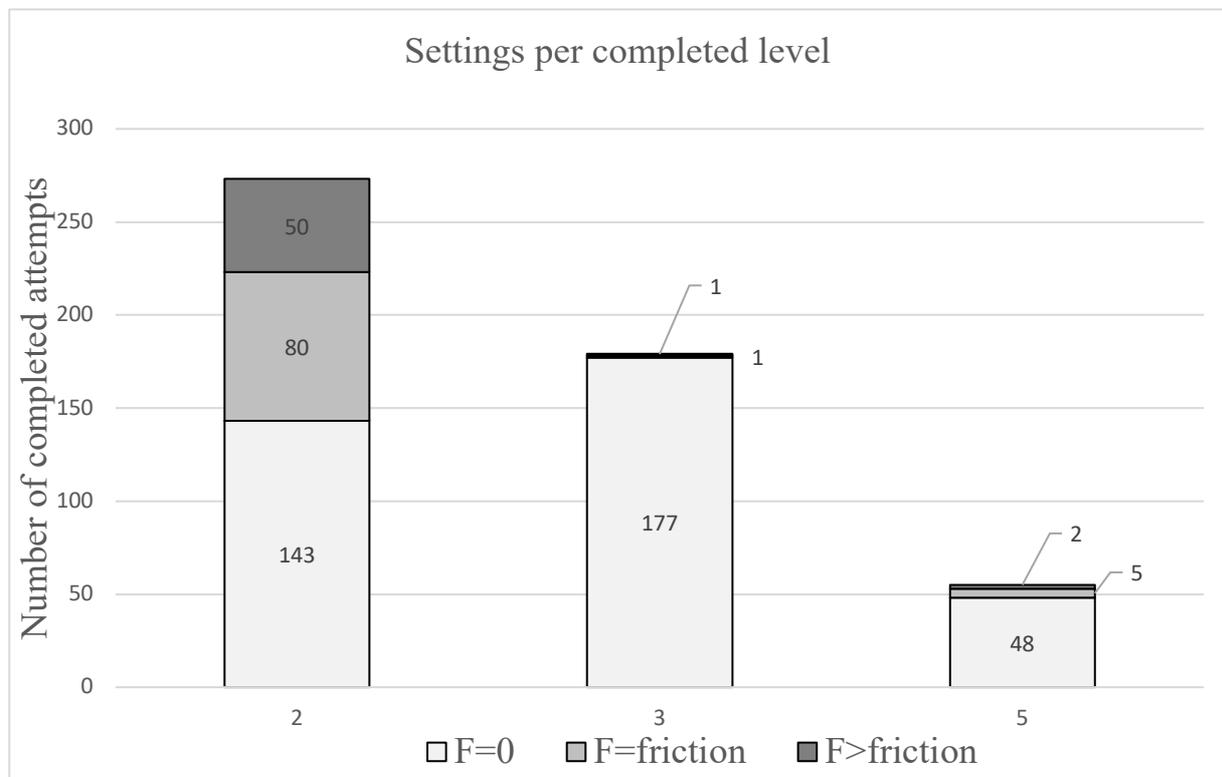

**Figure 7:** The number of completed attempts for each key level, with the chosen settings (F is equal to 0N (correct setting), F is equal to friction and F is bigger than friction).

Figure 7 gives an overview of the completed attempts of each key level. The key levels were completed 507 times, of which 368 times with the setting F = 0N. A chi-square test determined that this setting was chosen significantly more often than the other two settings; $\chi^2(1, N = 506) = 354.710$, $p < .000$. This indicates that in the completed attempts the correct setting, F = 0N, is most chosen.

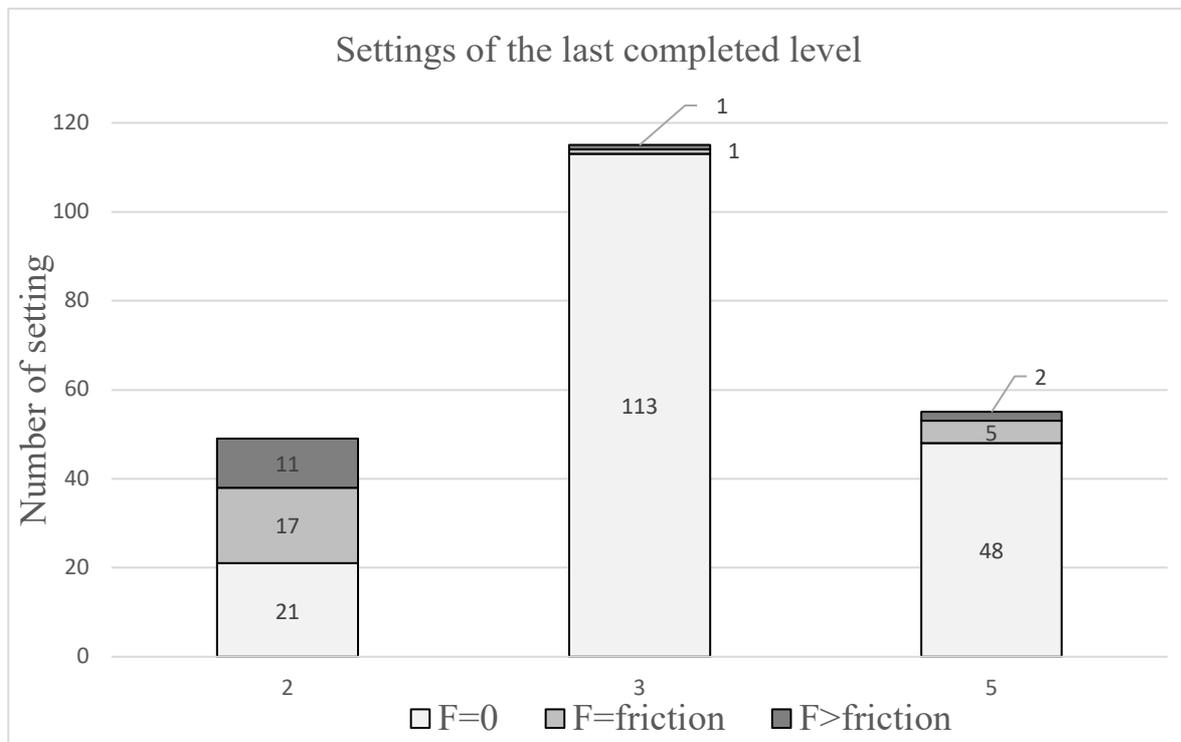

**Figure 8:** Each player had a last completed attempt in a key level. The number of the setting used for those attempts is displayed, F is equal to 0N (correct setting), F is equal to friction and F is bigger than friction.

For each player the setting of their last completed level was determined, as shown in Figure 8. An independent sample *t*-test was performed to examine the mean differences between the posttest scores of players who used the correct setting (M = 1.88, SD = 1.026) and players who did not (M = 1.86, SD = .948). No significant difference was found; $t(217) = .108$, $p = .914$. This indicates that players who used the incorrect setting to complete their last level learned as much from the game than players who used the correct setting to complete their last level.

4. **Conclusions and discussion**

**4.1 Conclusions**

The overarching research question was: *How does playing an intrinsically integrated game support learning of Newtonian mechanics?* In order to answer the overarching research question three sub-questions were composed.

SQ1: To what extent does participants' conceptual understanding of Newtonian mechanics change as a consequence of playing the game?

Results show a small ($d$ = .201) but significant learning effect after playing Newton's Race for 15 minutes. This indicates that players' conceptual understanding of Newtonian mechanics improved significantly in the relatively short intervention time.

SQ2: To what extent does the acquired knowledge transfer to different situations not closely related to the game?

The results show that players scored significantly less on questions outside the game context in comparison with questions closely related to the game. This indicates that players were only able to apply their knowledge to situations closely related to the game. They did not show the ability to transfer the acquired knowledge to other situations.

SQ3: To what extent does the participant's gameplay demonstrate evidence of intrinsic integration according to the guiding frame?

Results indicate a significant shift of the chosen setting as players progress through the game. The correct setting (F=0) is chosen significantly more often in the later levels of the game (Figure 6). As a notable exception, some players were able to finish levels with the incorrect scientific setting (Figures 7 & 8) demonstrating, at least for them, a discrepancy between the game goal and the learning goal (Figure 1). We may hypothesize these to be experienced gamers. Importantly however, the number of successful attempts with the incorrect setting (139) was much smaller than the number of successful attempts (368) with the correct setting. Furthermore, most successful level completions with incorrect setting occurred in the easier level 2 (130). In level 5 there were only 9 successful attempts with the incorrect setting. The fact that some players complete levels with incorrect settings raises the question if the players who ended their last completed level with the incorrect setting in fact did not reach the learning goal. However, results show no significant difference in learning effects between players who chose the incorrect and correct setting for their last completed level.

This brings us to the main research question: How does playing an intrinsically integrated game support learning of Newtonian mechanics?

We found players' conceptual understanding of Newtonian mechanics improved significantly after playing Newton's Race for 15 minutes. We surmise that they reached this learning effect by toying with the different settings, especially in key level 2, and opting for the correct setting more often as they progress through the levels. This is reflected in Figures 6,7, and 8 as the correct setting is chosen more often as the players progress. This indicates that the game mechanics support the chosen pedagogical approach: players are generally confronted with a problem (not finishing a level) if they opt for the incorrect setting (Figure 1). Figure 8 shows that most players finish their last level with the correct scientific setting, indicating that they knew that that was the one to use in order to complete the level. This may be seen as evidence for the idea that the game goal and the learning goal are aligned.

### 4.2 Situating the study

In educational game design, considering both pedagogical aspects and game aspects is relatively rare (Ávila-Pesántez et al., 2017). However, some recent studies show a more in depth analysis on the design of an educational game with the focus on combining gameplay with learning (Czauderna & Guardiola, 2019). The current study fits within the research body that emphasizes the importance of aligning gameplay with learning in educational games. Other studies either explicitly using intrinsic integration, or at least specifically aligning learning with gameplay, also found positive learning outcomes, (Habgood & Ainsworth, 2011; Czauderna & Guardiola, 2019). These studies also acknowledge the continuous investigation of the design process of such an educational game. The value of the present study is an attempt to find support for a guiding frame on intrinsic integration (Figure 1.) as a key element in the design process, by examining players' log data. Other studies have found a learning effect as well, however, we are able to enlighten the underlying mechanism by the results of the log data. These results strengthen the proposed guiding frame as a framework for designing an educational game.

### 4.3 Limitations and implications

In interpreting the results of the current study some limitations must be taken into account. First, participants were recruited on a voluntary basis which could indicate a positive attitude toward games in general from the participants. This could have an influence on the attitude from the participants towards Newton's Race. Also, the learning effect we found was quite small. However, we must take into consideration that participants demonstrated a significant increase in their conceptual understanding in just 15 minutes of gameplay, as no additional learning activities were used in this study. This shows promising possibilities for the use of Newton's Race in physics education. Finally, at this point we cannot conclude anything about a retention effect, since no delayed posttest was conducted. With this study we sought out to make a step in players conceptual understanding by playing Newton's Race. Research on a possible retention effect is a later step in the design process.

Results of this study show that an intrinsically integrated game on Newtonian mechanics can foster learning with just 15 minutes of gameplay. In order to extend the acquired knowledge on Newtonian mechanics and to transfer that knowledge onto situations outside the game, additional learning activities are needed. This finding is in line with the findings of Wouters and colleagues (2013). Our results could imply that similarly designed games on different (physics) subjects could also foster learning in a relative short time. A short educational game can thus be useful, because in a lesson there usually will the time to play the game and include additional learning activities.

### 4.4 Further research

For future research two main aspects must be taken into account. Firstly, it is important to embed the game in the curriculum to improve comprehension and transfer (Wouters, van Nimwegen, van Oostendorp & van der Spek, 2013). In such a lesson the game should play a central role and the other learning activities should complement the game. This lesson can probably best be given at the lower classes of secondary education, where students start with Newtonian mechanics. Also, findings indicate a positive correlation between age and posttest scores. In this study younger players participated in primary education and older players in the lower levels of secondary education. Secondly, the game itself needs some improvements. Observations indicate that players found the 'steering' counterintuitive. This is due to the fact that in most commercial games objects are steered the other way around. This could be an interesting point for an additional learning activity, such as a classroom discussion. Implementing these ideas in a future study could result in a fun and effective lesson which fosters comprehension and transfer on Newtonian mechanics.

**Declarations**

The present article was made possible by funding from the Dutch Ministry of Education, Culture and Science, OCW/PromoDoc/1065001. We would also like to thank Bas Bruil and Yannick Nales for their contribution to the programming of Newton's Race.
Conflict of Interest: The authors declare that thy have no conflict of interest.

**References**

Clark, D. B., Tanner-Smith, E. E., & Killingsworth, S. S. (2016). Digital games, design, and learning: A systematic review and meta-analysis. *Review of Educational Research*, *86*(1), 79–122.

Csíkszentmihályi, M. (1990). *Flow: The Psychology of Optimal Experience*. Harper & Row, New York.

Czauderna, A.; Guardiola, E. (2019). The Gameplay Loop Methodology as a Tool for Educational Game *Design. Electron. J. E-Learn.*, *17*, 207–221.

Denham, A. (2016). Improving the design of a learning game through intrinsic integration and playtesting. *Technology, Knowledge and Learning*, *21*(2), 175–194.

Driver, R., Squires, A., Rushworth, P., & Wood-Robinson, V. (1994). *Making Sense of Secondary Science: Research into children's ideas*. Routledge, Oxen.

Duit, R., & Treagust, D. (2003). Conceptual change: a powerful framework for improving science teaching and learning. *International Journal of Science Education*, *25*, 671-688.

Fazio, C. & Battaglia, O.R. (2018). Conceptual understanding of Newtonian mechanics through cluster analysis of FCI student answers. *International Journal of Science and Mathematics Education*, *17*, 1-21.

Habgood, M. P. J., & Ainsworth, S. E. (2011). Motivating children to learn effectively: exploring the value of intrinsic integration in educational games. *Journal of the Learning Sciences*, *20*(2), 169–206.

Halloun, I. A., & Hestenes, D. (1985). Common sense concepts about motion. *American Journal of Physics*, *53*(11), 1056.

Hestenes, D., Wells, M., & Swackhamer, G. (1992). Force concept inventory. *The physics teacher, 30*(3), 141–158.

Hewson, P. W., & Hewson, M. G. (1984). The role of conceptual conflict in conceptual change and the design of science instruction. *Instructional Science*, *13*, 1-13.

Kafai, Y. (1996). Learning design by making games: Children's development of strategies in the creation of a complex computational artifact. In Kafai, Y., Resnick, M. (eds.), *Constructionism in practice: Designing, thinking and learning in a digital world*. Erlbaum, Mahwah.

Ke, F. (2016). Designing and integrating purposeful learning in game play: a systematic review. *Educational Technology Research and Development*, *64*(2), 219–244.

Klaassen, K. (1995). *A problem-posing approach to teaching the topic of radioactivity*. Cdβ Press, Utrecht.

Kortland, J. (2001). *A problem posing approach to teaching decision-making about the waste issue*. Cdβ Press, Utrecht.

Lameras, P., Arnab, S., Dunwell, I., Stewart, C., Clarke, S., & Petridis, P. (2017). Essential features of serious games design in higher education: Linking learning attributes to game mechanics. *British Journal of Educational Technology*, *48*(4), 972–994.

Lijnse, P. L., & Klaassen, K. (2004). Didactical structures as an outcome of research on teaching learning sequences? *International Journal of Science Education, 26*(5), 537-554.

Van der Linden, A., & Van Joolingen, W. R. (2016). A serious game for interactive teaching of Newton's laws. *Proceedings of the 3rd Asia-Europe Symposium on Simulation and*

*Serious Gaming - 15th ACM SIGGRAPH Conference on Virtual-Reality Continuum and its Applications in Industry, VRCAI 2016*, 165–167

Van der Linden, A., Van Joolingen, W. R., & Meulenbroeks, R. F. G. (2019). Designing an intrinsically integrated educational game on Newtonian mechanics. *International Conference on Games and Learning Alliance, 11385*, 123–133.

Ávila-Pesántez, D., Rivera, L., Alban, M. (2017). Approaches for Serious Game Design: A Systematic Literature Review. *Computers in Education Journal*, *8*(3), 1–11.

Schumacher, R. S., Hofer, S., Rubin, H., & Stern, E. (2016). How Teachers Can Boost Conceptual Understanding in Physics Classes. In Looi, C. K., Polman, J. L., Cress, U., Reimann, P. (eds.), *Transforming Learning, Empowering Learners: The international Conference of the Learning Sciences*: *Vol. 2* (pp. 1167-1168). International Society of the Learning Sciences, Singapore.

Sicart, M. (2008). Designing game mechanics. *International Journal of Computer Game Research, 8*(2).

Vandercruysse, S., & Elen, J. (2017). Towards a game-based learning instructional design model focusing on integration. In Wouters, P., van Oostendorp, H. (eds.), *Instructional Techniques to Facilitate Learning and Motivation of Serious Games*. Springer International Publishing Switzerland, Cham.

Vollebregt, M.J. (1998). *A problem posing approach to teaching an initial particle model*. Cdβ Press, Utrecht

Vosniadou, S. (1994) Capturing and modeling the process of conceptual change. *Learning and Instruction*, *4*, 45-69.

Wouters, P., van Nimwegen, C., van Oostendorp, H., & van der Spek, E. D. (2013). A meta-analysis of the cognitive and motivational effects of serious games. *Journal of Educational Psychology, 105*(2), 249–265.

Zeng, J., Parks, S., & Shang, J. (2020). To learn scientifically, effectively, and enjoyably: A review of educational games. *Human Behavior and Emerging Technologies, 2*(2).

**Appendix A: Pretest (translated from original language)**

1. If you kick a stationary ball, the ball will start moving.
True / False / I do not know

2. If the ball starts moving after a kick, there will be a force working on the ball in the direction of motion. True / False / I do not know

3. You kick a ball through long grass. When you stop kicking, the ball will stop moving immediately. True / False / I do not know

4. You kick a ball once, so that it rolls over grass. The ball will stop moving eventually. True / False / I do not know

5. You kick a ball once, so that it rolls over grass. Does a force work on the ball in the direction of motion after the kick?
After the kick:
A. There is no force working.
B. There is a force working. The force is equal to opposing forces (from the grass).
C. There is a force working. The force is bigger than opposing forces (from the grass).
D. I do not know.

6. To let a ball roll across a soccer field (grass), you must kick the ball 10 times. If this soccer field was made of ice, would you have to kick the ball less, as many times of more often?
A. Less
B. As many times
C. More often
D. I do not know.

7. You kick a ball once, so that it will roll across grass. What kind of motion will the ball make?
A. First faster and then slower.
B. Ever slower.

C. First slower and then faster.

D. Ever faster.

E. A constant speed.

F. I do not know.

8. You kick a ball across grass, so that the ball moves with a constant speed. The forwards kick-force on the ball is:

A. Bigger than opposing forces (from for instance the grass).

B. Smaller than opposing forces (from for instance the grass).

C. Equal to opposing forces (from for instance the grass).

D. I do not know.

**Appendix B: Posttest (translated from original language)**

1. If you kick a stationary ball, the ball will start moving. After the kick, there will be a force working on the ball in the direction of motion.
True / False / I do not know

2. You kick a ball through long grass. When you stop kicking, the ball will stop moving immediately. True / False / I do not know

3. You roll a hockey ball through long grass. When you stop pushing, the ball will stop moving immediately. True / False / I do not know

4. You kick a ball once, so that it rolls over grass. Does a force work on the ball in the direction of motion after the kick?
After the kick:
A. There is no force working.
B. There is a force working. The force is equal to opposing forces (from the grass).
C. There is a force working. The force is bigger than opposing forces (from the grass).
D. I do not know.

5. You kick a ball once, so that it will roll across grass. What kind of motion will the ball make?
A. First faster and then slower.
B. Ever slower.
C. First slower and then faster.
D. Ever faster.
E. A constant speed.
F. I do not know.

6. You kick a ball across grass, so that the ball moves with a constant speed. The forwards kick-force on the ball is:
A. Bigger than opposing forces (from for instance the grass).
B. Smaller than opposing forces (from for instance the grass).
C. Equal to opposing forces (from for instance the grass).

D. I do not know.

7. You roll a hockey ball across grass, so that the ball moves with a constant speed. The forwards roll-force on the ball is:

A. Bigger than opposing forces (from for instance the grass).

B. Smaller than opposing forces (from for instance the grass).

C. Equal to opposing forces (from for instance the grass).

D. I do not know.